\documentclass[a4paper,fleqn,usenatbib,useAMS]{mnras}
\usepackage{newtxtext,newtxmath}
\usepackage[T1]{fontenc}
\usepackage{ae,aecompl}
\usepackage{graphicx}    
\usepackage{amsmath}     
\usepackage{amssymb}     
\usepackage{multicol}    
\usepackage{bm}          
\usepackage{pdflscape}   
\usepackage{xspace}
\newcommand{\Msun}{\,$M_{\sun}$\xspace}
\newcommand{\Rsun}{\,$R_{\sun}$\xspace}
\newcommand{\kms}{\,km\,s$^{-1}$\xspace}
\title[The puzzle of SN~2016X] 
   {Resolving the puzzle of type IIP SN~2016X}
\author[V. P. Utrobin \& N. N. Chugai]{
V. P. Utrobin$^{1,2}$\thanks{E-mail: utrobin@itep.ru}
and
N. N. Chugai$^{3}$
\\
$^{1}$NRC ``Kurchatov Institute'' --
      Institute for Theoretical and Experimental Physics,
      B.~Cheremushkinskaya St. 25, 117218 Moscow, Russia \\
$^{2}$Max-Planck-Institut f\"ur Astrophysik, Karl-Schwarzschild-Str. 1,
      85748 Garching, Germany \\
$^{3}$Institute of Astronomy, Russian Academy of Sciences, Pyatnitskaya
      St. 48, 119017 Moscow, Russia
}

\date{Accepted XXX. Received YYY; in original form ZZZ}

\pubyear{2019}

\begin{document}
\label{firstpage}
\pagerange{\pageref{firstpage}--\pageref{lastpage}}
\maketitle
%
\begin{abstract}
The enigmatic type IIP SN 2016X demonstrates the unprecedented asphericity in
   the nebular H$\alpha$ line profile, the absence of nebular [\ion{O}{i}]
   emission, and the unusual occultation effect due to the internal dust.
The hydrodynamic modelling of the bolometric light curve and expansion 
   velocities suggests that the event is an outcome of the massive star
   explosion that ejected 28\Msun with the kinetic energy of
   $1.7\times10^{51}$\,erg and 0.03\Msun of radioactive $^{56}$Ni.
We recover the bipolar distribution of $^{56}$Ni from the H$\alpha$ profile
   via the simulation of the emissivity produced by non-spherical $^{56}$Ni
   ejecta.
The conspicuous effect of the dust absorption in the H$\alpha$ profile rules
   out the occultation by the dusty sphere or dusty thick disk but turns out
   consistent with the thin dusty disk-like structure in the plane
   perpendicular to the bipolar axis.
We speculate that the absence of the nebular [\ion{O}{i}] emission might
   originate from the significant cooling of the oxygen-rich matter mediated
   by CO and SiO molecules.
\end{abstract}
\begin{keywords}
hydrodynamics -- methods: numerical -- supernovae: general --
supernovae: individual: SN 2016X
\end{keywords}

\section{Introduction} 
\label{sec:intro}
Type IIP supernovae (SNe~IIP, ``P'' stands for plateau at the light curve)
   originate from massive stars that retain a significant fraction of the
   hydrogen envelope until the core collapse.
A pre-SN star before the explosion has the structure of a red supergiant (RSG)
   \citep{GIN_71, Sma_09} which favors the high luminosity at the plateau
   powered by the explosion energy.
The plateau with a duration of about 100\,days is followed by the radioactive
   tail which, in turn, is powered by $^{56}$Co decay.
SNe~IIP show a broad range of plateau luminosities: from
   $\sim$2$\times10^{41}$\,erg\,s$^{-1}$ for subluminous SNe~IIP, e.g.,
   SN~2003Z \citep{UCP_07}, to $\sim$10$^{43}$\,erg\,s$^{-1}$ for SN~2009kf,
   the most energetic case among ordinary SNe~IIP \citep{UCB_10}.
SN~1987A-like supernovae related to the explosion of a blue supergiant (BSG)
   sometimes are classified as peculiar SNe~IIP, although their broad luminosity
   maximum at about 90--100\,days is completely powered by the radioactive
   decay of $^{56}$Co \citep{Woo_88}.

According to the evolutionary models, SNe~IIP originate from the progenitors
   in the range of $9 - M_\mathrm{\,IIP}$ with $M_\mathrm{\,IIP} \approx
   25$\Msun for solar metallicity and $M_\mathrm{\,IIP} \approx 40$\Msun for
   low metallicities \citep{HFWLH_03}.
The hydrodynamic modelling of the well-observed SNe~IIP recovers the ejecta
   masses in the range of 13.1--28.1\Msun for the sample of 10 SNe~IIP
   \citep{UC_17}.
This is in line with the ejecta mass range of 12.4--25.6\Msun for the sample
   of 9 SNe~IIP, inferred by \citet{Nad_03} using the scaling relations based
   on the hydrodynamic models \citep{LN_85}. 
These ``hydrodynamic'' ejecta masses combined with the neutron star mass lead
   to the minimum progenitor masses in the range of 14--30\Msun without taking
   into account the mass loss.
A comparison of the archive photometry of SN~IIP progenitors with the
   evolutionary RSG models indicates the lower progenitor masses of 8--18\Msun
   \citep{Sma_15}, that brings about yet unsettled tension between the
   hydrodynamic masses and the masses recovered from the archival photometry
   \citep[but see e.g.][]{PZS_17}.

There is a general consensus that the SN~IIP explosion is caused by the core
   collapse, although the conversion of the binding energy of a newly born
   neutron star into the kinetic energy of the ejecta is not yet fully
   understood.
The preferred mechanism is the neutrino-driven explosion, however the successful
   self-consistent model is available so far only for low energy
   ($\approx$10$^{50}$\,erg) events related to $\approx$10\Msun progenitors
   \citep{Jan_17}. 
The general physical arguments admit that the neutrino-driven explosion is able
   to provide the energy of up to $2\times10^{51}$\,erg \citep{Jan_17}.
The alternative mechanism is the magneto-rotational explosion driven by either
   magnetic bipolar jets \citep{LW_70, WMW_02, BMA_18}, or by amplified
   toroidal field along the equatorial plane \citep{Bis_71, BMA_18}.
The intrinsic feature of the magneto-rotational explosion seems to be a bipolar
   ejecta asymmetry.
Another remarkable property of the magneto-rotational mechanism is its
   potential to produce the high-energy events \citep{BDL_07} that could
   account for SN~2009kf with the explosion energy of $2.15\times10^{52}$\,erg
   \citep{UCB_10}.

The magneto-rotational explosion of SNe~IIP thus might be indicated by
   the high explosion energy and the ejecta asymmetry.
The explosion energy of SNe~IIP along with the ejecta mass and the pre-SN
   radius can be recovered only via the modelling of the light curve and
   expansion velocities of the well-observed objects.
As to the explosion asymmetry, it could be imprinted in the asphericity of
   the $^{56}$Ni ejecta.
The latter, in turn, can be revealed via the H$\alpha$ asymmetry at the nebular
   stage when the line emissivity closely traces the energy deposition of
   gamma-rays and positrons from the $^{56}$Co decay \citep{Chu_07}.
   
Until recently the most conspicuous H$\alpha$ asymmetry in SNe~IIP was
   demonstrated by SN~2004dj \citep{CFS_05} and was interpreted as an outcome
   of the bipolar $^{56}$Ni ejecta.
Less pronounced, yet apparent, is the asymmetry of SN~2013ej \citep{MDJ_17}
   that shows signatures of the asymmetric high-velocity $^{56}$Ni
   ejecta \citep{UC_17}.
The asymmetry shown by the recent type IIP SN~2016X \citep{BDER_19}
   significantly outclasses that of SN~2004dj.
The nebular H$\alpha$ profile looks weird: two separated peaks of comparable
   intensity with the deep minimum in between.
This H$\alpha$ profile unambiguously indicates the bipolar $^{56}$Ni ejecta
   \citep{BDER_19} presumably observed along the bipolar axis.

But this is not the only surprise demonstrated by SN~2016X.
Even more unusual is the absence of the SN~IIP generic [\ion{O}{i}] 6300,
   6364\,\AA\ nebular emission.
This emission is barely seen on day 340 and completely absent at day 471
   \citep{BDER_19}.
At the moment this puzzling phenomenon remains a challenging problem.

Finally, we couldn't help noticing the unusual occultation effect due
   to the internal dust.
On day 471 the red H$\alpha$ peak is significantly attenuated by the dust
   which signals the internal dust formation \citep{BDER_19} at the right
   time for SNe~IIP.
On day 740 the red H$\alpha$ peak completely disappears which reflects
   the increase of the amount of the internal dust.
The surprising fact however is that the blue H$\alpha$ peak does not show
   any sign of the additional blueshift caused by the occultation, which
   would be present in the case of the central (quasi-)spherical dusty zone
   likewise in SN~1987A \citep{LDGB_89, MIW_17} and SN~1999em
   \citep{ECP_03}.

The unusual manifestations of SN~2016X raise a question, whether we see
   the explosion of a normal massive RSG with the progenitor mass and
   the explosion energy typical for explored SNe~IIP, i.e., $M \sim 13-30$\Msun
   and $E \sim (0.2-2)\times10^{51}$\,erg \citep{UC_17}, or we face
   an extraordinary  event.
Fortunately, photometric and spectral observations \citep{HWH_18, BDER_19}
   provide us with an excellent basis to explore different aspects of
   SN 2016X in detail and possibly to clear up the issue.

Below SN 2016X will be studied in several ways.
We start with the hydrodynamic modelling to determine principal SN parameters,
   i.e., the ejecta mass, the kinetic energy, and the pre-SN radius;
   the amount of $^{56}$Ni will be obtained directly from the luminosity in
   the radioactive tail.
We then recover the $^{56}$Ni distribution in the envelope from the nebular
   double-peaked H$\alpha$ by means of the computation of the energy
   deposition produced by the asymmetric $^{56}$Ni distribution.
This simulation includes effects of the absorption of the H$\alpha$ emission
   by the dust, which will permit us to constrain the spatial distribution
   of the dusty material.
Finally, we will use X-ray observations during the first 20 days after
   the explosion \citep{GDS_16, BDER_19} to infer the pre-SN wind density and
   to test a compatibility of the SN hydrodynamic parameters with the
   observational effects of the ejecta/wind interaction.

Throughout the paper we use the distance $D = 15.2\pm3.0$\,Mpc \citep{BDER_19}
   and the reddening $E(B-V) = 0.04$\,mag \citep{HWH_18}.
The explosion date is set to be 2016 January 18.7 that is recovered from
   the fit of the earliest $V$ magnitudes by the hydrodynamic model.
This moment is only 0.25\,days earlier compared to that adopted by
   \citet{HWH_18}.

\section{Hydrodynamic modelling}
\label{sec:hydro}
The principal SN parameters, i.e., the explosion energy, the ejecta mass, and
   the pre-SN radius are recovered via the hydrodynamic description of the
   observational bolometric light curve and the expansion velocity at
   the photosphere; the latter should be preferably fitted at the early
   photospheric stage, when the effects of the internal asymmetry of the
   radioactive $^{56}$Ni ejecta are minimal.
Another crucial parameter, the total amount of $^{56}$Ni, is directly inferred
   from the flux at the early (<200\.days) stage of the radioactive tail.
It is reasonable to start the section with clarifying the data that constitute
   the observational bases for the hydrodynamic modelling of SN~2016X.

\subsection{Observational data}
\label{sec:hydro-odata}
The available $UBVRI$ photometry \citep{HWH_18} is used to recover the 
   bolometric light curve of SN~2016X.
The filter fluxes are fitted by a black body spectrum multiplied by the
   reduction factor that takes into account the spectrum suppression in
   the blue and ultraviolet regions primarily due to the line opacity.
The reduction factor depends on the SN age and its wavelength
   dependence is recovered from the SN~1987A spectral energy distribution
   at different ages \citep{PKS_95}.
The photometric data taken by the 0.8-m Tsinghua University-NAOC telescope (TNT)
   and the Lijiang 2.4-m telescope (LJT) \citep{HWH_18} are in good
   agreement and used to recover the observational bolometric light curve.

The time evolution of the observed photospheric velocity is taken from
   \citet{HWH_18}.
In addition to the photospheric velocity, we need the maximal velocity of
   the SN ejecta because of its key role in constraining hydrodynamic model.
The point is that the maximal velocity is sensitive to the choice of the
   radius of the pre-SN model and its density structure.
The earliest spectrum that provides us with a reliable value of the maximal
   velocity is the spectrum at 4.56\,days.
Our estimate of the velocity from the blue edge of the H$\alpha$ emission at
   that moment is $13750\pm500$\kms, in accordance with the value reported
   by \citet[Fig.~10]{HWH_18}.

\subsection{Model overview}
\label{sec:hydro-model}
%
\begin{figure}
   \includegraphics[width=\columnwidth, clip, trim=0 237 54 138]{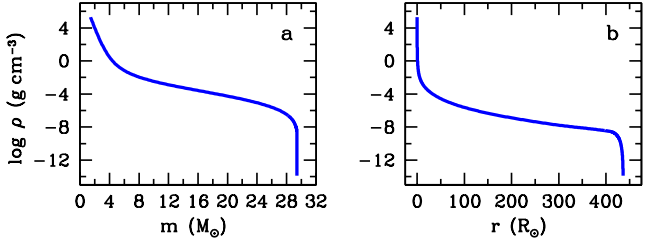}
   \caption{%
   Density distribution as a function of interior mass (Panel \textbf{a}) and
      radius (Panel \textbf{b}) for the pre-SN model.
   The central core of 1.4\Msun is omitted.
   }
   \label{fig:denmr}
\end{figure}
\begin{figure}
   \includegraphics[width=\columnwidth, clip, trim=8 17 46 250]{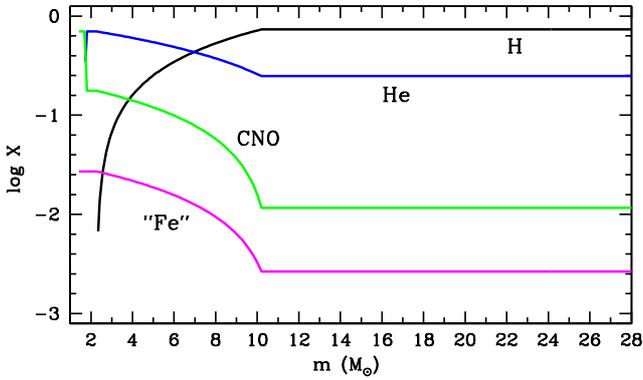}
   \caption{%
   Mass fraction of hydrogen (\emph{black line\/}), helium
      (\emph{blue line\/}), CNO elements (\emph{green line\/}),
      and Fe-peak elements excluding radioactive $^{56}$Ni
      (\emph{magenta line\/}) in the ejected envelope.
   }
   \label{fig:chcom}
\end{figure}
The one-dimensional hydrodynamic code with the radiation transfer
   \citep{Utr_04} is used to explode the hydrostatic non-evolutionary
   pre-SN model.
This approach is preferred because the pre-SN model produced by the
   stellar evolution computations is generally unable to describe SN~IIP
   observations.
This was understood in the wake of SN~1987A \citep{Woo_88}.
A serious modification of the pre-SN model is required that includes
   the extended mixing between the metal-rich ejecta, the He-rich and H-rich
   envelopes with the smoothing of steep density gradients.
The hand-made mixing is used therefore for the pre-SN model in order to
   imitate in the one-dimensional model the mixing produced by the real
   essentially three-dimensional explosion \citep{UWJM_17}.

The acceptable pre-SN model is found via numerical simulations of a set of
   models with different SN parameters.
The explosion of pre-SN model is initiated by a supersonic piston applied to
   the stellar envelope at the boundary with the collapsing 1.4\Msun core.
The pre-SN density and chemical composition of the optimal model are shown
   in Fig.~\ref{fig:denmr} and Fig.~\ref{fig:chcom}, respectively.
We did not solve the optimization problem rigorously, since this procedure
   requires enormous computational efforts.
Instead, the optimal model is recovered as a compromise between the fits to
   the observed light curve and the evolution of the velocity at the
   photosphere.

\subsection{Optimal model and supernova parameters}
\label{sec:hydro-modpar}
%
\begin{figure}
   \includegraphics[width=\columnwidth, clip, trim=0 16 54 245]{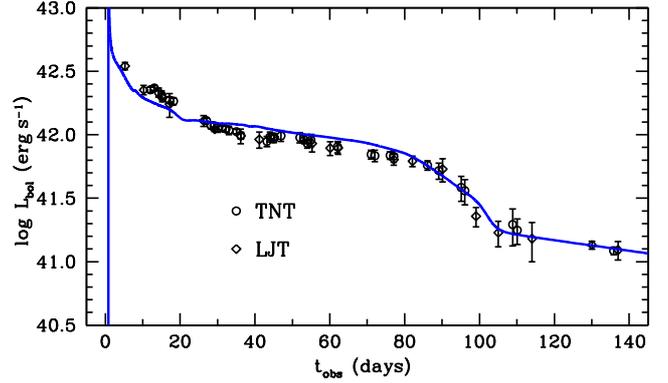}
   \caption{%
   The model bolometric light curve (\emph{blue line\/}) overplotted on
      the bolometric luminosities which we recovered from the photometric data 
      reported by \citet{HWH_18}.
   }
   \label{fig:blc}
\end{figure}
\begin{figure}
   \includegraphics[width=\columnwidth, clip, trim=6 16 54 245]{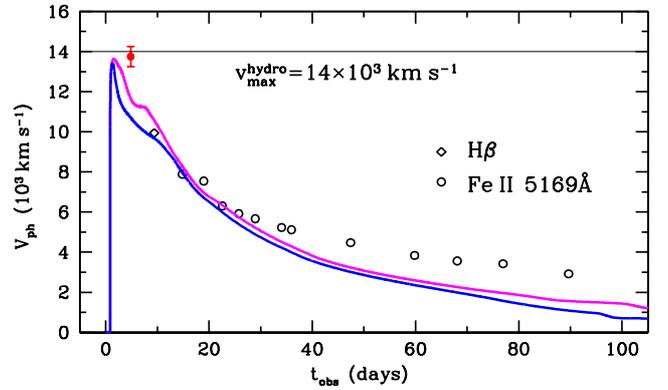}
   \caption{%
   The evolution of model velocity at the photosphere defined by the level
      $\tau_{eff} = 2/3$ (\emph{blue line\/}) and $\tau_\mathrm{T} = 2/3$
      (\emph{magenta line\/}) is compared with the photospheric velocities
      estimated from the line absorption minimum \citep{HWH_18}.
   The red point is the early expansion velocity obtained from the blue
      wing of H$\alpha$.
   The horizontal line is the model maximal velocity corresponding to
      the strongest density peak formed at the shock breakout stage;
      the interaction with the wind is neglected.
   For the late-time mismatch see Section~\ref{sec:hydro-modpar}.
   }
   \label{fig:vph}
\end{figure}
\begin{figure}
   \includegraphics[width=\columnwidth, clip, trim=15 18 54 213]{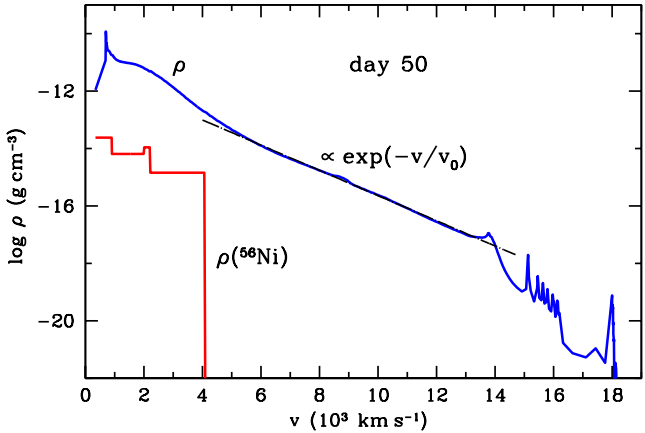}
   \caption{%
   The gas (\emph{blue line\/}) and $^{56}$Ni (\emph{red line\/}) density
      as a function of velocity at $t=50$\,days.
   \emph{Dash-dotted line\/} is the exponential fit $\exp(-v/v_0)$.
   }
   \label{fig:denni}
\end{figure}
\begin{figure}
   \includegraphics[width=\columnwidth, clip, trim=0 18 52 244]{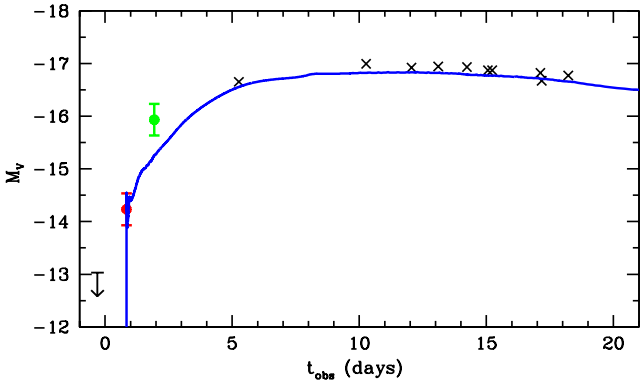}
   \caption[]{%
   $V$-band light curve (\emph{blue line\/}) during the first 20 days for
      the optimal model.
   Arrow marks the upper limit $V>18.0$\,mag for a non-detection on
      January 18.351, red point is the average of the two earliest $V$
      estimates separated by 0.008\,days with the uncertainty of 0.3\,mag,
      and green point is the $V$ magnitude of 15.1 at January 20.586
      \citep{BSS_16}.
   Crosses are the observational points of \citet{HWH_18}.
   The model explosion date is January 18.7.
   }
   \label{fig:onset}
\end{figure}

The optimal hydrodynamic model satisfactorily reproduces the bolometric light
    curve (Fig.~\ref{fig:blc}) and the expansion velocity at the early stage
   (Fig.~\ref{fig:vph}).
The model maximal velocity specified by the density peak at 14000\kms
   (Fig.~\ref{fig:denni}) is also consistent with the observed maximal velocity
   of $13750\pm500$\kms recovered from the H$\alpha$ emission at 4.56\,days.
Note that both definitions of the photosphere via the effective and
   Thomson opacity predict close velocity values.
The model $V$-band light curve fits satisfactorily the initial
   behavior of the absolute $V$ magnitude including the discovery point
   (Fig.~\ref{fig:onset}); the shown fit suggests that the explosion occurred
   at 2016 January 18.7, i.e., 0.25\,days earlier compared to the explosion
   moment adopted by \citet{HWH_18}. 

\begin{figure}
   \includegraphics[width=\columnwidth, clip, trim=51 16 24 188]{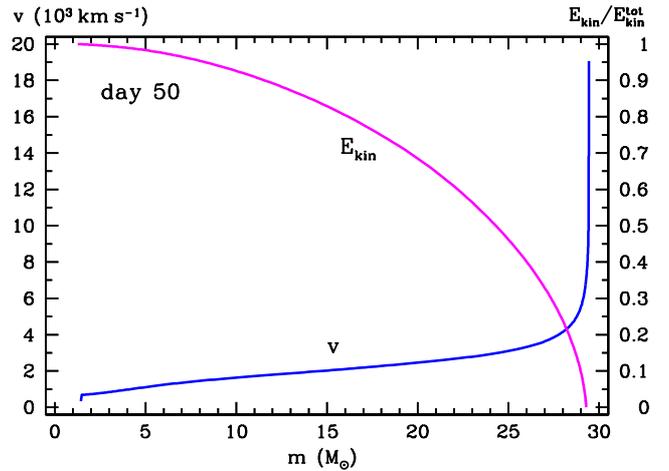}
   \caption{%
   Distribution of velocity and kinetic energy along the mass coordinate
      in the SN ejecta of the optimal model on day 50.
   }
   \label{fig:mev}
\end{figure}

The optimal model is specified by the ejecta mass $M = 28$\Msun, the
   kinetic energy $E = 1.73\times10^{51}$\,erg, and the pre-SN radius
   $R_0 = 436$\Rsun.
The $^{56}$Ni mass directly recovered from the radioactive tail is 0.0295\Msun.
The radial distribution of $^{56}$Ni in the model is a spherical representation
   of the bipolar $^{56}$Ni ejecta recovered from the modelling of the
   H$\alpha$ profile at the nebular stage (Section~\ref{sec:ha}).
The total density and the density of $^{56}$Ni in the freely expanding envelope
   is shown in Fig.~\ref{fig:denni}.
The oscillatory structure of the density distribution in the outermost layers
   ($v > 14000$\kms) forms at the shock breakout and is related to the
   instability of the radiative acceleration due to the strong opacity
   dependence on the temperature and density.
The characteristic property of the optimal model is the large fraction of the
   kinetic energy residing in the outer layers: the 4\Msun external ejecta,
   about 14\% of the total ejecta mass, contain about 50\% of the total kinetic
   energy (Fig.~\ref{fig:mev}).
This feature is closely related to the ability of the hydrodynamic model to
   reproduce both the initial luminosity peak and the high expansion velocity
   of the external layers.

The uncertainty in the derived SN parameters can be estimated by a variation
   of the model parameters around the optimal model.
The uncertainty of the distance (see Section~\ref{sec:intro}) implies
   the 40\% uncertainty in the bolometric luminosity. 
The scatter in the plot of the photospheric velocity versus time
   (Fig.~\ref{fig:vph}) suggests the uncertainty of 7\% in the photospheric
   velocity. 
We estimate the maximal uncertainty of the plateau length as 3\,days, i.e.,
   3\% of the plateau duration. 
With these uncertainties of observables, we find the errors of
   $\pm360$\Rsun for the initial radius, $\pm2.1$\Msun for the ejecta
   mass, $\pm0.19\times10^{51}$\,erg for the explosion energy, and
   $\pm0.012$\Msun for the total $^{56}$Ni mass.

The model reveals some deviations from the data, which require comments.
The luminosity excess at the plateau (Fig.~\ref{fig:blc}) stems from the
   spherical approximation of the bipolar $^{56}$Ni distribution.
Given the bipolar geometry of the $^{56}$Ni ejecta, the escaping flux should
   be anisotropic at the late photospheric stage when the radioactivity
   contributes to the escaping luminosity.
Furthermore, according to the H$\alpha$ model (Section~\ref{sec:ha}),
   the rear $^{56}$Ni component has larger both mass and velocity compared
   to the front component which implies that the backside photosphere
   is brighter than the front one.
The observed ``isotropic'' luminosity ($4 \pi D^2 f$) defined via the observed
   flux $f$ thus underestimates the overall SN~2016X luminosity, which accounts
   for the model flux excess at the plateau.
Another mismatch is the lower velocity at the photosphere compared to the
   observed values after about 30\,days (Fig.~\ref{fig:vph}).
This disparity also stems from the spherical approximation of the $^{56}$Ni
   distribution.
Indeed, the bipolar $^{56}$Ni ejecta result in the prolate shape of the
   photosphere with the large axis aligned along the line of sight.
The observed velocities of absorption minima at the late plateau therefore
   are larger than the photospheric velocities of the spherical model.
 
\subsection{Significance of early stage}
\label{sec:hydro-estage}
The determination of SN~IIP parameters is based on describing the bolometric
   or monochromatic light curves and the evolution of expansion velocity at
   the photospheric level by means of hydrodynamic modelling.
It is obvious that the derived parameters are more reliable in the case of
   a SN~IIP well observed photometrically and spectroscopically from the
   explosion moment till the radioactive tail.
Here we would like to emphasize a significant role of the initial
   ($t < 30$\,days) stage in recovering the SNe~IIP parameters, because
   this issue is oftentimes missed.

The issue was partially explored for the well-observed SN~2005cs \citep{UC_08}.
The optimal hydrodynamic model in this case is characterized by the parameter
   set $M_{ej} = 15.9$\Msun, $E = 4.1\times10^{50}$\,erg, and $R_0 = 600$\Rsun.
On the other hand, ignoring the fit to the initial stage permits us to describe
   the plateau of the light curve and the evolution of the photospheric velocity
   at the ages $t > 30$\,days by the explosion of a 9\Msun pre-SN star with
   the parameters $M_{ej} = 7.8$\Msun, $E = 1.4\times10^{50}$\,erg,
   and $R_0 = 700$\Rsun \citep{UC_08}.
Neglecting the early stage in the hydrodynamic modelling for SN~2005cs
   thus leads to a strong disagreement with the optimal model.
A similar mismatch between the models with the full and reduced approaches was
   recently demonstrated for SN~1999em \citep[see][Fig.~13]{UWJM_17}.

An alternative ``easy-to-use'' approach to estimate the basic SN~IIP parameters
   is provided by the Litvinova-Nadyozhin relations between the parameters and
   the SN observables: the plateau duration, the luminosity and the
   photospheric velocity at the middle of the plateau \citep{LN_85}.
Although these relations are based on the extended grid of hydrodynamic models
   one should keep in mind that the models are not aimed at the description
   of the full data set on the bolometric light curve and the expansion
   velocities, so the parameters derived with this approach may differ by
   a significant factor from those determined via the hydrodynamic modelling
   of a particular SN~IIP.
Another drawback of this approach is neglecting the influence of radioactive
   $^{56}$Ni on the light curve.
The hydrodynamic parameters of the well-observed SN~1999em are
   $M_{ej} = 19$\Msun, $E = 1.3\times10^{51}$\,erg, and $R_0 = 500$\Rsun
   \citep{Utr_07}, whereas the Litvinova-Nadyozhin relations result in
   $M_{ej} = 15$\Msun, $E = 0.68\times10^{51}$\,erg, and $R_0 = 414$\Rsun
   \citep{Nad_03}, i.e., 20\% lower ejecta mass and twice as low explosion
   energy.
For SN~2016X the Litvinova-Nadyozhin relations suggest $M_{ej} = 23.5$\Msun,
   $E = 1.76\times10^{51}$\,erg, and $R_0 = 130$\Rsun, i.e., 16\% lower ejecta
   mass and 3 times lower pre-SN radius compared to our optimal model.

To summarize, neglecting the description of the light curve and expansion
   velocities at the early epoch, $t < 30$\,days, can significantly
   affect the SN~IIP parameters inferred via the radiation hydrodynamic
   modelling of the RSG explosion in the framework of setting described above.

\section{Double-peaked H$\alpha$ and dust distribution}
\label{sec:ha}
The double-peaked H$\alpha$ profile in late nebular spectra of SN~2016X
   \citep{BDER_19} is attributed to the bipolar $^{56}$Ni ejecta embedded
   in the spherical hydrogen envelope.
The striking feature of SN~2016X is that the H$\alpha$ peaks are fully
   separated by the deep minimum \citep[Fig.~1]{BDER_19} which indicates
   that we look at the supernova almost along the bipolar axis.
On days 471 and 740 the H$\alpha$ is affected by the dust absorption:
   the red peak first gets weaker and on day 740 completely disappears.
The model for the H$\alpha$ at late stages therefore should include absorption
   by the internal dust.
 
The central to our model is the assumption that the bipolar $^{56}$Ni ejecta
   do not disturb the overall hydrogen spherical symmetry.
For the accepted bipolar $^{56}$Ni distribution the gamma-ray energy
   deposition is calculated in the single flight approximation.
The effective absorption coefficient for gamma-quanta of $^{56}$Co decay
   is approximated as $k_{\gamma} = 0.06 Y_e$\,cm$^2$\,g$^{-1}$, where
   $Y_{e}$ is a number of electrons per nucleon \citep{KF_92}.
Positrons of $e^-$-capture deposit their kinetic energy on-the-spot.
The H$\alpha$ emissivity is assumed to be proportional to the local deposition
   rate; the emissivity saturation due to the complete ionization never
   attains at the relevant epochs. 

The additional hydrogen ionization by the photoionization from
   the second level is neglected; this process however dominates at
   the early (195\,days) nebular stage when the peaks contrast is
   relatively small and on day 142 when the double-peaked structure is not
   seen at all \citep{HWH_18}.
The transformation between days 142 and 195 is related to the significant
   decrease of the rate of the hydrogen photoionization from the second level
   compared to the non-thermal ionization and excitation.
On day 340 the photoionization from the second level is negligible, so the
   H$\alpha$ emissivity rate is uniquely linked with the local deposition
   of the energy of gamma-quanta and positrons of $^{56}$Co decay,
   which favours the reliable inference of the $^{56}$Ni distribution from
   the H$\alpha$ profile.
The hydrogen abundance is assumed to be solar and homogeneous all over
   envelope except for the central region $v < v_h = 500$\kms where
   no hydrogen is assumed.
The density distribution of the homologously expanding envelope is exponential,
   $\rho = \rho_0\exp{(-v/v_0)}$ with $\rho_0\propto t^{-3}$ and $v_0$
   determined by the ejecta mass of 28\Msun and the kinetic energy of
   $1.73\times10^{51}$\,erg. 

\begin{table}
\centering 
\caption[]{Parameters of $^{56}$Ni components.}
\label{tab:ni}
\begin{tabular}{lccc}
\hline
\noalign{\smallskip}
Component & $v_s$ & $v_r$  & $\mu$  \\
          & \multicolumn{2}{c}{(km\,s$^{-1}$)} &  \\
\noalign{\smallskip}
\hline
\noalign{\smallskip}
 front   & 1100  &  1100   &  1    \\
 rear    & 2400  &  1600   &  1.5  \\
 central &  0    &  2000   &  0.075 \\
\noalign{\smallskip}
\hline
\end{tabular}
\end{table}
\begin{figure}
   \includegraphics[width=\columnwidth, clip, trim=52 112 71 127]{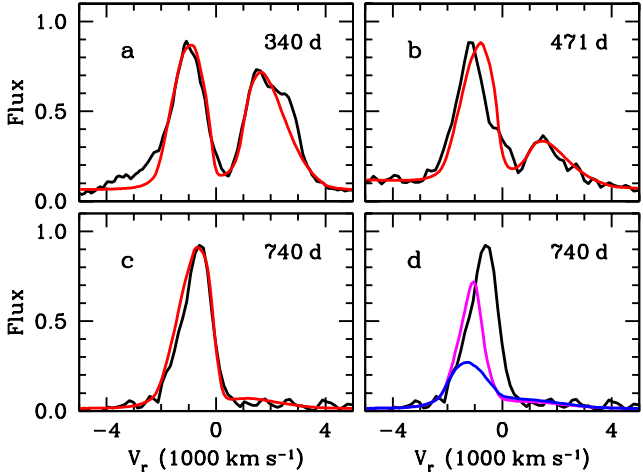}
   \caption{%
   H$\alpha$ line in nebular spectra of SN~2016X (\emph{black line\/}) with
      the overplotted profile of model (\emph{red line\/}) in
      Table~\ref{tab:ni} on days 340, 471, and 740.
   Panel \textbf{a}: on day 340 the dust absorption is absent.
   Panel \textbf{b}: the spectrum on day 471 is reproduced by the same model
      as on day 340 that includes a geometrically thin dusty disk with
      the optical depth $\tau_d = 0.85$ shifted by 100\kms towards far side
      (see Fig.~\ref{fig:cart}).
   Panel \textbf{c}: the spectrum on day 740 with the same model as on day 471,
      but for the dust optical depth of 3.
   Panel \textbf{d}: the spectrum on day 740 cannot be reproduced with the
      dusty sphere (\emph{blue line\/}) nor the geometrically thick dusty
      disk (\emph{magenta line\/}).
   }
   \label{fig:ha}
\end{figure}
The bipolar $^{56}$Ni distribution is represented by the front and rear
   homogeneous spheres.
We also tried homogeneous ellipsoids and conies but with less success.
The central spherical component with the boundary velocity of 2000\kms
   is also included.
All the components lie on the same axis arbitrarily inclined by
   $i = 10^{\circ}$ with respect to the line of sight; in fact the observed
   profile admits the inclination angle of the bi-polar axis in the range of
   $i \lesssim 20^{\circ}$.
Table~\ref{tab:ni} contains the derived shift ($v_s$), radius ($v_r$),
   and the relative mass of components for the optimal model
   (Fig.~\ref{fig:ha}). 
Interestingly, the recovered bipolar structure is asymmetric: the mass and
   the shift velocity of the rear component are significantly larger compared
   to the front component (Table~\ref{tab:ni}).

\begin{figure}
   \includegraphics[width=0.9\columnwidth, clip, trim=202 271 79 144]{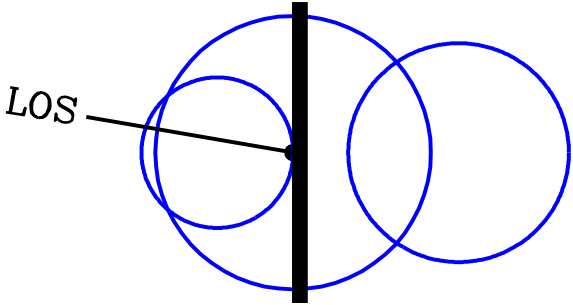}
   \centering
   \caption{%
   The configuration of the three homogeneous spheres of $^{56}$Ni in the
      model of double-peaked H$\alpha$.
   The black vertical line section renders the dusty disk in the model of
      H$\alpha$ on day 740.
   Slanted line section shows the line of sight (LOS) assuming $10^{\circ}$
      inclination.
   }
   \label{fig:cart}
\end{figure}
On days 471 and 740 the H$\alpha$ is strongly affected by the dust formed in
   the inner ejecta \citep{BDER_19}.
Remarkably, an attempt to describe this effect in terms of the central dusty
   sphere, likewise it has been done in the case of SN~1987A \citep{LDGB_89}
   and SN~1999em \citep{ECP_03}, fails (Fig.~\ref{fig:ha}d).
The dusty thick disk aligned perpendicular to the bipolar axis with the
   diameter/thickness ratio of 5 is also ruled out (Fig.~\ref{fig:ha}d).
While the red component can be fully absorbed for both dust configurations,
   the blue peak is modified by the dust absorption to an extent that makes
   both models unacceptable.
The best fit on day 740 (Fig.~\ref{fig:ha}c) is attained in the model with
   a thin dusty disk of the radius of 2200\kms and the optical depth
   $\tau_d = 3$, aligned perpendicular to the bipolar axis and shifted by
   $\approx$100\kms towards rear component.
It should be emphasized, that the model circular plane disk is an idealization.
In reality this could be a non-circular irregular disk-like structure.
On day 471 the same model requires the disk optical depth of 0.85
   (Fig.~\ref{fig:ha}b).
The overall configuration of $^{56}$Ni components with the model dusty disk
   is shown in Fig.~\ref{fig:cart}.

\begin{table}
\centering 
\caption[]{Parameters of \ion{Ca}{ii} components.}
\label{tab:ca}
\begin{tabular}{lccc}
\hline
\noalign{\smallskip}
Component & $v_s$ & $v_r$  & $\mu$  \\
          & \multicolumn{2}{c}{(km\,s$^{-1}$)} &  \\
\noalign{\smallskip}
\hline
\noalign{\smallskip}
 front   &  800  &   800   &  1    \\
 rear    & 2400  &  1600   &  0.2  \\
 central &    0  &  2000   &  1.25 \\
\noalign{\smallskip}
\hline
\end{tabular}
\end{table}
\begin{figure}
   \includegraphics[width=0.9\columnwidth, clip, trim=124 130 80 220]{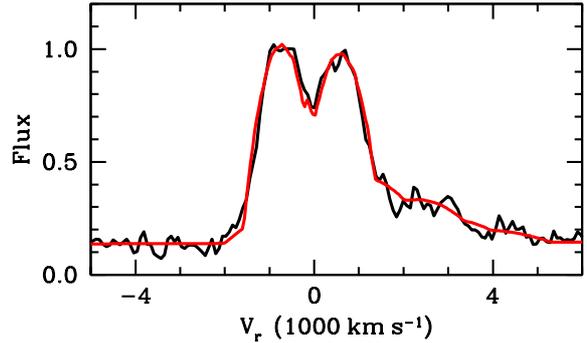}
   \centering
   \caption{%
   [\ion{Ca}{ii}]~7291, 7324\,\AA\ doublet in SN~2016X on day 340
      (\emph{black line\/}) with the overplotted model profile
      (\emph{red line\/}) (Table \ref{tab:ca}) that demonstrates the
      dominance of the front component.
   Zero of radial velocity corresponds to the rest frame position of
      the 7281.47\,\AA\ line.    
   }
   \label{fig:ca}
\end{figure}
It should be emphasized that the absence of the dust extinction on day 340
   is consistent with the fact that the dust in SN~1987A and SN~1999em forms
   only after 400\,days.
This however raises a question, what is the origin of the blueshift of emission
   peaks of [\ion{Ca}{ii}]~7291, 7324\,\AA\ doublet reported by \citet{BDER_19}.
This cannot be the effect of the internal dust because even at the early
   nebular phase (195\,days) the [\ion{Ca}{ii}] doublet shows the similar
   asymmetry.
We suggest that the line asymmetry in the [\ion{Ca}{ii}] doublet is related to
   the asymmetry of the luminosity of the front and rear bi-polar components
   of [\ion{Ca}{ii}] emission.
This possibility is illustrated in Fig.~\ref{fig:ca} that shows [\ion{Ca}{ii}]
   doublet on day 340 with the overplotted synthetic profile.
The model includes three spherical components of homogeneous emissivity similar
   to the $^{56}$Ni distribution.
Models parameters, i.e., shift, radius, and relative luminosities of
   \ion{Ca}{ii} components are given in Table~\ref{tab:ca}.
These values emphasize the fact that the doublet luminosity of the front
   component is 5 times larger than rear one.
This asymmetry can arise both from the different Ca masses or/and different
   ionization and excitation conditions in the front and rear components.
The electron number densities in the components are comparable so, if the
   temperatures are also comparable, then the front component
   contains several times larger amount of Ca compared to the rear component.
It might well be that the asymmetry of Ca components is the another
   manifestation of the asymmetry of $^{56}$Ni components.

The model for [\ion{Ca}{ii}] profile includes additional parameter, the
   ratio $R$ of blue-to-red emissivity that, in turn, depends on the
   line Sobolev optical depth.
We find that the observed profile requires $R = 1.3$ (compared to $R = 1.5$
   for the optically thin case).
The recovered ratio corresponds to the optical depth in the 7291\,\AA\ line
   $\tau(7291\mbox{\AA}) = 1.08$.
Since the \ion{Ca}{ii} luminosity of the front component dominates and its
   radius is minimal, the \ion{Ca}{ii} density is maximal in the front
   component.
The recovered optical depth therefore refers primarily to the front component.
The Sobolev optical depth can be converted to the \ion{Ca}{ii} mass of
   $1.1\times10^{-3}$\Msun for a given volume of the front component.
Assuming a linear scaling between the mass and luminosity we obtain the
   total \ion{Ca}{ii} mass of three components of
   $\approx2.7\times10^{-3}$\Msun.
Given a possible contribution of \ion{Ca}{iii}, (\ion{Ca}{ii} can be easily
   ionized by Ly$\alpha$ quanta) the found mass of Ca should be considered
   as the lower limit. 

\section{Wind density and X-rays}
\label{sec:wind}
The reported X-ray luminosity of SN 2016X \citep{BDER_19} can be used to
   recover the density of the circumstellar (CS) gas lost by the pre-SN.
To this end we employ the interaction model based on the thin shell
   approximation \citep{Che_82, Giu_82}.
The model was described earlier \citep{CCL_04} and we recap here only essential
   points.
The gas swept up by the forward and reverse shock forms the shell whose
   expansion rate is governed by the equations of motion and mass conservation.
The X-ray luminosity of both shocks at the moment $t$ is calculated as
   the shock kinetic luminosity with the factor of the radiation efficiency
   $\eta = t/(t+t_c)$, where $t_c$ is the cooling time of the postshock gas
   calculated for the density four times of the preshock density.
The SN density distribution is set to be exponential
   $\rho = \rho_0\exp{(-v/v_0)}$, which is in line with the hydrodynamic
   model (cf. Fig.~\ref{fig:denni}).
Parameters $\rho_0$ and $v_0$ are specified by the ejecta mass $M_{ej}$ and
   the kinetic energy $E$.
The escaping X-rays are subject to the absorption in the SN ejecta and in the
   cool dense shell that forms due to the cooling of shocked ejecta in the
   reverse shock.
To compare the model X-ray luminosity with the observed values, we take into
   account only the X-ray radiation in the range $h\nu < 10$\,keV in
   accordance with the reported {\em Chandra} and {\em Swift} data
   \citep{BDER_19}.
The X-ray emission from the reverse shock gets into this band, while the
   relatively low contribution of the forward shock luminosity is taken
   into account adopting the spectrum $\epsilon^{-0.5}\exp{(-\epsilon/kT)}$.

\begin{figure}
   \includegraphics[width=\columnwidth, clip, trim=52 132 10 198]{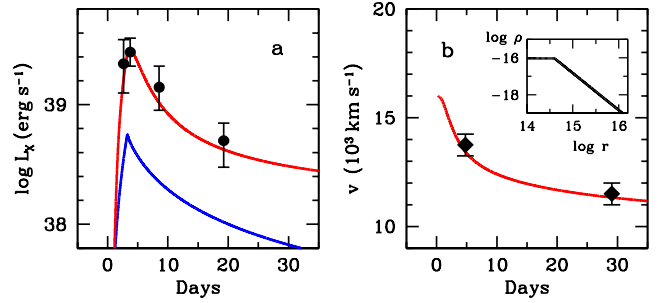}
   \caption{%
   The X-ray luminosity in the range of $< 10$\,keV and the boundary velocity
      of the unshocked ejecta for the CS interaction model.
   Panel \textbf{a}: the total escaping X-ray luminosity (\emph{red line\/})
      with contribution of the forward shock luminosity (\emph{blue line\/}).
   Dots are the {\em Chandra} and {\em Swift} luminosity in $0.3-10$\,keV band
      \citep{BDER_19}.
   Panel \textbf{b}: the model boundary velocity of the unshocked ejecta
      (\emph{red line\/}).
   Diamonds show the maximal velocity of the ejecta recovered from
      the H$\alpha$ blue emission wing on day 4.6 and from the blue edge of
      the H$\alpha$ absorption component at 28.8\,days in the spectrum of
      \citet{HWH_18}.
   Inset shows the CS density distribution.
   }
   \label{fig:xray}
\end{figure}
The model X-ray luminosity reproduces the reported data (Fig.~\ref{fig:xray}a)
   for the ejecta parameters $M_{ej} = 28$\Msun, $E = 2\times10^{51}$\,erg, and
   the CS density distribution $\rho = const$ in the range of
   $r < 4\times10^{14}$\,cm and $\rho \propto r^{-2}$ for larger radii
   (Fig.~\ref{fig:xray}b, inset).
The model boundary velocity of the unshocked ejecta well fits the velocity
   at 4.56\,days found from the blue wing of the H$\alpha$ emission and
   the velocity at 28.8\,days found from the blue edge of the H$\alpha$
   absorption component (Fig.~\ref{fig:xray}b).
To summarize, the SN and CSM models reproduce both the X-ray data and the
   evolution of the maximal velocity of the unshocked ejecta.
Yet it should be emphasized that the recovered CS density distribution can be
   also consistent with the other options of $M_{ej}$ and $E$, provided their
   values obey the scaling $E \propto M_{ej}^{\,0.88}$.

The found wind density at $r > 4\times10^{14}$\,cm is characterized by the
   parameter $w = \dot{M}/u = 1.9\times10^{14}$\,g\,cm$^{-1}$.
It is useful to express this parameter via convenient units as
   $w = \dot{M}_{-6}/u_{10} = 3$, where $\dot{M}_{-6}$ is in units of
   $10^{-6}$\Msun\,yr$^{-1}$ and $u_{10}$ is in units of 10\kms.
The inferred value $w = 3$ is three times as large as that for the type IIP
   SN~1999em and SN~2004dj \citep{CCU_07}.
Since the mass-loss rate increases with the stellar mass one can conclude
   that the progenitor of SN~2016X was more massive compared to both mentioned
   SNe~IIP.
In order to find the RSG mass loss rate, one needs to know the wind velocity.
The wind velocity of Milky Way RSGs with a mass of $\sim$30\Msun 
   (e.g. $\mu$\,Cep and VX\,Sgr) is 20\kms \citep{MJ_11}.
Assuming the same wind velocity for SN 2016X, the mass-loss rate of its
   progenitor turns out to be $6\times10^{-6}$\Msun\,yr$^{-1}$.
 
\section{Discussion and Conclusions}
\label{sec:disc}
\begin{figure}
   \includegraphics[width=\columnwidth, clip, trim=0 23 27 27]{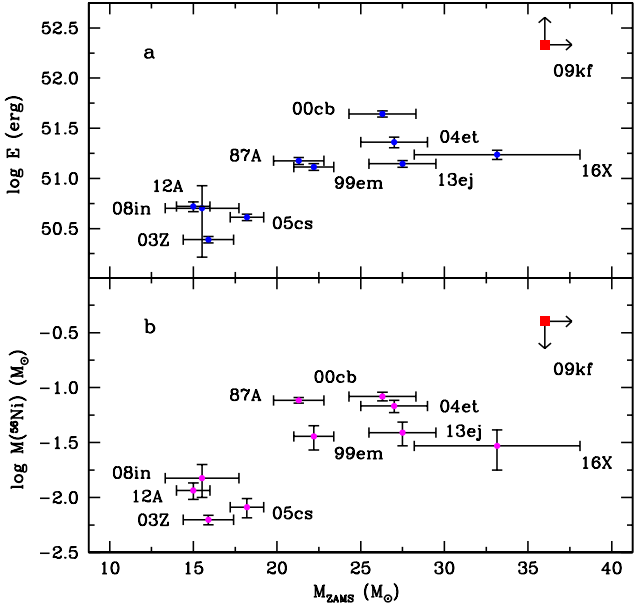}
   \caption{%
   Explosion energy (Panel \textbf{a}) and $^{56}$Ni mass (Panel \textbf{b})
      vs. hydrodynamic progenitor mass for SN~2016X and ten other
      core-collapse SNe \citep{UC_17}.
   }
   \label{fig:ennims}
\end{figure}

%
\begin{table}
\centering
\caption[]{Hydrodynamic models of type IIP supernovae.}
\label{tab:sumtab}
\begin{tabular}{@{ } l  c  c @{ } c @{ } c @{ } c  c @{ }}
\hline
\noalign{\smallskip}
 SN & $R_0$ & $M_{ej}$ & $E$ & $M_{\mathrm{Ni}}$ 
       & $v_{\mathrm{Ni}}^{max}$ & $v_{\mathrm{H}}^{min}$ \\
       & (\Rsun) & (\Msun) & ($10^{51}$\,erg) & ($10^{-2}$\Msun)
       & \multicolumn{2}{c}{(km\,s$^{-1}$)}\\
\noalign{\smallskip}
\hline
\noalign{\smallskip}
 1987A &  35  & 18   & 1.5    & 7.65 &  3000 & 600 \\
1999em & 500  & 19   & 1.3    & 3.6  &  660  & 700 \\
2000cb &  35  & 22.3 & 4.4    & 8.3  &  8400 & 440 \\
 2003Z & 230  & 14   & 0.245  & 0.63 &  535  & 360 \\
2004et & 1500 & 22.9 & 2.3    & 6.8  &  1000 & 300 \\
2005cs & 600  & 15.9 & 0.41   & 0.82 &  610  & 300 \\
2008in & 570  & 13.6 & 0.505  & 1.5  &  770  & 490 \\
2009kf & 2000 & 28.1 & 21.5   & 40.0 &  7700 & 410 \\
2012A  &  715 & 13.1 & 0.525  & 1.16 &  710  & 400 \\
2013ej & 1500 & 26.1 & 1.4    & 3.9  &  6500 & 800 \\
 2016X &  436 & 28.0 & 1.73   & 2.95 &  4000 & 760 \\
\noalign{\smallskip}
\hline
\end{tabular}
\end{table}
The recovered parameters of peculiar SN~2016X suggest rather high ejecta mass
   compared to other SNe~IIP (Table~\ref{tab:sumtab}) which were studied
   earlier on and whose parameters were summarized by \citet{UC_17}.
Allowing for the collapsed core, the pre-SN mass amounts to 29.4\Msun
   which should be considered as a lower limit for the main-sequence mass
   of the progenitor.
The mass lost by the BSG wind during the main-sequence phase can be estimated
   unfortunately only with a large uncertainty.
The theoretical bolometric luminosity of a 30\Msun progenitor
   is $\approx$4$\times10^5\,L_{\sun}$ \citep[e.g.][]{EFSMC_13}.
Observational mass-loss rate for a BSG of that luminosity lies in the range
   from $5\times10^{-8}$ to $\sim$10$^{-6}$\Msun\,yr$^{-1}$
   \citep[cf.][]{KK_17}.
Given the lifetime at the hydrogen burning stage of about $6\times10^6$\,yr
   \citep{MMSSC_94}, the lost mass thus turns out in the range of $\sim$0.3 to
   $\sim$6\Msun.
At the RSG stage, the recovered mass-loss rate of about
   $6\times10^{-6}$\Msun\,yr$^{-1}$ and the RSG lifetime of
   $\approx$10$^5$\,yr \citep{MCE_15} result in the lost mass of $\sim$0.6\Msun.
The total mass lost by the progenitor thus lies in the range of 0.9 to
   6.6\Msun.
These values combined with the error of the ejecta mass are translated into
   the SN~2016X progenitor mass range from 28.2 to 38.1\Msun.
Even with the minimal progenitor mass of about 30\Msun, SN~2016X turns out
   the most massive among normal SNe~IIP on the scatter diagrams
   $E$ vs. $M_\mathrm{ZAMS}$ and $M$($^{56}$Ni) vs. $M_\mathrm{ZAMS}$
   (Fig.~\ref{fig:ennims}).
We speculate that the large progenitor mass could be somehow related to  
   peculiar manifestations of SN~2016X including the bipolar structure of
   the $^{56}$Ni ejecta, the low luminosity of the [\ion{O}{i}] doublet,
   and the unusual disk-like distribution of the dusty material.

The pronounced bipolar $^{56}$Ni ejecta is likely produced by the bipolar
   explosion asymmetry.
Factors that favored this asymmetry could include a large scale instability
   at the core collapse or/and a rotation.
It is likely that the dusty disk-like structure is also an outcome the bipolar
   asymmetry.
This structure could be related to either the dense two-dimensional
   condensation in the equatorial plane formed during the explosion, or
   the fragment of the dense shell of the $^{56}$Ni bubble in the far
   hemisphere.

The absence of the normal [\ion{O}{i}] emission might be explained by
   the low amount of the synthesized oxygen in the ejected envelope.
There could be two reasons for that: (i) the low-mass progenitor 
   $M_\mathrm{ZAMS} \approx 10$\Msun with the pre-SN devoid of the oxygen
   mantle around the collapsing core, or (ii) the fallback of the oxygen
   shell onto the black hole. 
The first possibility should be discarded since it contradicts to the large
   ejecta mass.
The second option cannot be ruled out because we are not aware of the explosion
   details.
  
An alternative possibility is that the O-rich matter in SN~2016X ejecta
   is too cold for the normal [\ion{O}{i}] emission.
This situation could arise, if CO and SiO molecules are formed at the nebular
   stage all over the oxygen ejecta.
The conjecture follows the findings that the cooling of the O-rich matter
   via CO and SiO molecules in SN~1987A strongly suppresses the nebular
   [\ion{O}{i}] emission, so the observed [\ion{O}{i}] emission comes out only
   from the oxygen matter devoid of molecules \citep{LD_95}.
For SN~2016X the luminosity of the [\ion{O}{i}] 6300, 6364\,\AA\ on day 340
   is $\leq$7.7$\times10^{37}$\,erg\,s$^{-1}$ according to the spectrum
   reported by \cite{BDER_19}.
The oxygen-core mass of a 30\Msun progenitor is of 8\Msun \citep{WHW_02}.
Taking into account the collapsed 1.4\Msun core and assuming the solar
   C/O ratio, we obtain the 4.6\Msun oxygen ejecta.
The luminosity of the [\ion{O}{i}] doublet from this amount of oxygen meets
   the observation constraint, if the excitation temperature of the oxygen
   is $\leq$2000\,K.
This requirement is easily fulfilled since the temperature in the oxygen zone
   of SN~1987A cooled by CO and SiO molecules is about 1800\,K during the
   first year and later on gets lower \citep{LD_95}.
The SN~2016X is presumably a special case in which CO and SiO molecules form
   throughout the O-rich matter.
This assumption combined with a moderate amount of $^{56}$Ni would result in
   the strong cooling of all the O-rich gas thus inhibiting [\ion{O}{i}]
   emission.
One may conjecture that the required Si and O abundance in the oxygen-rich
   matter is the outcome of the He and C burning in combination with the
   convection and rotation-induced mixing \citep[e.g.][]{Heg_98}.
If the efficient CO and SiO cooling occurs in SN~2016X, one expects that
   similar high-mass SNe~IIP at the nebular stage should demonstrate a strong CO
   and SiO emission in the vibrational fundamental and first overtone bands.

The recovered mass of the SN~2016X ejecta aggravates the well-known disparity
   between the relatively high masses of SN~IIP progenitors estimated by means
   of the hydrodynamic modelling of the well-observed SNe~IIP
   (Table~\ref{tab:sumtab}) and the lower progenitor masses inferred from
   the archival photometry using the stellar evolution models.
At the moment the stellar evolution theory is unable to reliably fix the upper
   boundary of the mass range producing SNe~IIP, $M_\mathrm{\,IIP}$, because
   the resulting loss of the hydrogen envelope is a matter of the adopted
   prescription for mass loss and rotation effects.
Recent evolutionary calculations of massive stars suggest that
   $M_\mathrm{\,IIP} \sim 20$\Msun at the solar metallicity without rotation
   \citep{LC_18}.
This is lower than the former estimate of $\sim$30--35\Msun \citep{LC_10}.
The reassessment is caused by a high mass-loss rate taken from \citet{LCZL_05}.
However, one should keep in mind that this mass-loss prescription suffers from
   large uncertainties.
Particularly, it predicts a larger mass-loss rate by a factor of about 1.8\,dex
   compared to that of the well-studied Galactic RSG $\alpha$\,Ori and
   $\mu$\,Cep \citep[Figure~11]{LCZL_05}.
The situation with the value of $M_\mathrm{\,IIP}$ predicted by the theory of
   stellar evolution thus looks rather uncertain with the conservative estimate
   of $M_\mathrm{\,IIP}$ in the range of 20--35\Msun for stars with the solar
   metallicity and zero rotation velocity.

\section*{Acknowledgements}
We thank Subo Dong for kindly sharing spectra of SN~2016X.
V.P.U. is partially supported by Russian Scientific Foundation grant
   19-12-00229.


\bsp	
\label{lastpage}
\end{document}